# Introgression makes waves in inferred histories of effective population size

John Hawks

Revision March 27, 2017

Affiliation:

Department of Anthropology

University of Wisconsin-Madison

email: jhawks@wisc.edu

## Abstract

Human populations have a complex history of introgression and of changing population size. Human genetic variation has been affected by both these processes, so that inference of past population size depends upon the pattern of gene flow and introgression among past populations. One remarkable aspect of human population history as inferred from genetics is a consistent "wave" of larger effective population size, prior to the bottlenecks and expansions of the last 100,000 years. Here I carry out a series of simulations to investigate how introgression and gene flow from genetically divergent ancestral populations affect the inference of ancestral effective population size. Both introgression and gene flow from an extinct, genetically divergent population consistently produce a wave in the history of inferred effective population size. The time and amplitude of the wave reflect the time of origin of the genetically divergent ancestral populations and the strength of introgression or gene flow. These results demonstrate that even small fractions of introgression or gene flow from ancient populations may have large effects on the inference of effective population size.

## Introduction

The origins of today's modern human populations included introgression or gene flow from genetically divergent ancestral populations of archaic humans. Direct evidence of this ancient introgression comes from the ancient DNA of Neandertals, which contributed between 1 and 4 percent of the genetic makeup of today's populations outside Africa (Green et al., 2010; Prüfer et al., 2013) and from the Denisova 3 genome, which represents an archaic population that contributed up to 5 percent of the genetic makeup of today's Melanesian and Australian populations (Reich et al., 2010; Reich et al., 2011; Meyer et al., 2012; Prüfer et al., 2013). No ancient DNA evidence from any archaic human skeletal remains within Africa has yet been recovered. However, abundant indirect evidence exists of introgression from genetically divergent populations into the ancestral populations of living Africans. Genetic comparisons extending to whole genomes now show the signature of archaic human introgression in samples of various populations within Africa (Lachance et al., 2012; Hsieh et al., 2016b; Beltrame et al., 2016).

At the same time, whole-genome analyses have added a longer-term dimension to our understanding of human population size over time (Li and Durbin, 2011; Mallick et al., 2016). But the inference of changes in effective population size depends in part upon the same features of genetic data that allow the inference of introgression. The genetic variation within a population is a product not only of its size but also its structure, and many aspects of human population history may complicate the understanding of effective population size (Hawks, 2008). While it is possible to explicitly employ complex population history models to generate estimates of past population size and structure (e.g., Schaffner et al., 2005), our ability to probe evidence of extinct "ghost" populations that contributed introgression or gene flow into living people is extremely limited.

Some studies have examined how demic expansion may have interacted with archaic human introgression during the expansion of modern human populations into Eurasia (Currat and Excoffier, 2011; Sugden and Ramachandran, 2016). However, the effects of introgression upon signatures of ancient population growth that predate the dispersal of modern humans into Eurasia have not previously been subject to close examination. Hence, it is valuable to look directly at how a history of introgression or gene flow between populations may affect the inference of ancestral effective population size.

**History of human effective population size.** Statistical methods such as the pairwise sequential Markovian coalescent (PSMC) (Li and Durbin, 2011) make it possible to infer changes in effective population size over time based upon the time to most recent common ancestor (TMRCA) of alleles of single genetic loci, extended across many loci up to whole genomes. The TMRCA indirectly reflects the genealogical coalescence of alleles, conditioned on the probability of mutations and recombination in each generation. Under a Wright-Fisher population model, the probability of coalescence of two alleles in a single generation is $1/2N_e$, and if the effective size in such a population changed in the past, different intervals of time will have different probabilities of coalescence. Li and Durbin (2011) showed the effectiveness of this approach to inferring changes in past population size as applied to single diploid genomes. They found that although the recent population history of human groups differs, for the time intervals prior to 100,000 years ago, different human genomes give repeatable and consistent estimates of effective population size.

These approaches have been applied to many samples of living humans as well as to ancient human genomes (e.g., Li and Durbin, 2011; Prüfer et al., 2013; Mallick et al., 2016). Several aspects of human population histories as inferred from the PSMC approach are notable. Non-African population samples reflect a bottlenecked history in which the ancestral effective size was rather strongly reduced between 100,000 and 50,000 years ago. Some sub-Saharan African populations also show a moderate reduction in effective size across that time interval, which may reflect either their own population history or admixture from populations of Eurasian origin, possibly via North Africa (Mallick et al., 2016). Sub-Saharan hunter-gatherer populations do not present any evidence of such a population bottleneck prior to 50,000 years ago, although their current population sizes are relatively small. All living human population samples in the period before 100,000 years ago show a similar inferred pattern of changes in effective population size. In broad terms looking backward into the past, this population history resembles a "wave" in which the effective size was large around 100,000—200,000 years ago, gradually smaller further back in time, with a minimum between 500,000 and 1 million years ago, and then gradually larger in the period before a million years ago (Figure 1). The inference of larger ancestral effective population size prior to 1 million years ago is repeated across different populations. It is also evident when considering genomes from archaic humans, specifically high-coverage Neandertal and Denisovan genomes from Denisova Cave, Russia (Prüfer et al., 2013). These genomes reflect an extremely small ancestral effective size through most of the evolutionary history of these groups back to their differentiation from other human populations before 500,000 years ago. But in the interval preceding their divergence, both samples share an inferred history of effective population size similar to that of living human populations. The inference of higher effective population size in the earliest part of human origins is also evident from rare genomic insertions (Huff et al., 2010). Both the replication in ancient genomes with very different recent population histories, and the evidence for a similar expansion from non-SNP datasets providing corroborating evidence that the appearance of a "wave" in effective population size is not merely an artifact of the PSMC method.

**History of introgression.** Statistical evidence for introgression comes from joint consideration of local heterozygosity and recombination in genetic samples. Recombination and mutation are both random processes that occur over time, leading to the expectation that regions of high local heterozygosity (because of a long history of mutations) should tend to be relatively short (because of an equally long history of recombination). If a population contains two long alleles that differ by a relatively large number of mutations, this suggests that some process may have suppressed recombination during the evolution of the locus. Relatively long, strongly linked divergent haplotypes can occur within a random-mixing population for several reasons, including balancing selection or structural variation such as chromosomal inversion (Evans et al., 2006; Hawks et al., 2008). However, if such loci are statistically common including across regions without notable selective or structural constraints, introgression of alleles from a genetically divergent population is a likely explanation.

Plagnol and Wall (2006) developed an approach to test the hypothesis of introgression of DNA into a population from a genetically diverged ancestral population, using the statistic $S*$. This method relies upon the idea that for a genetic locus with some proportion of introgressed DNA sequences, these sequences will tend to be distinguished by a relatively large number of single nucleotide variants that tend to be in linkage disequilibrium. Such a pattern repeated at many loci will appear inconsistent with random mating and therefore suggest some proportion of descent from one or more genetically divergent populations. Using this approach, Plagnol and Wall (2006) found evidence for introgression at a level of approximately 5% in multilocus data from European and from Yoruba samples, suggesting for the first time that intermixture with extinct populations had contributed substantially to the ancestry of African peoples as well as Eurasians. Significantly, the result preceded the recovery of ancient chromosomal DNA from Neandertal, Denisovan and early modern human specimens, which would demonstrate introgression based on direct genome comparisons (Green et al., 2010; Reich et al., 2010; Prüfer et al., 2013; Fu et al., 2015).

In adapted form, the *S\** approach has been applied by a series of investigators to African population samples. Hammer and colleagues (2011) found evidence for ancient introgression within samples of Biaka and San peoples, consistent with a model of 2% admixture from a genetically divergent population 35,000 years ago. They further found some candidate loci for introgression had widespread geographic distributions across Africa. Using whole-genome data from Western Pygmy, Hadza, and Sandawe individuals, Lachance et al. (2012) supported the hypothesis of introgression from an extinct population or populations with genetic divergence as great as the Neandertals. Some candidate loci for introgression were private within each population; others were shared across all three populations. Hsieh et al. (2016b) examined whole genome sequences from Baka and Biaka individuals, finding evidence for introgression from one or more genetically diverged populations extending over a possibly long period of time, with a pulse as recently as 9000 years ago.

Statistical modeling of introgression as applied to African population samples suggests an origin of some genetically divergent source populations for introgression around the time that Neandertals diverged from the ancestors of living African populations (Lachance et al., 2012), which we now think was prior to 700,000 years ago (Meyer et al., 2016). If these genetically divergent archaic populations originated from a common ancestral population around this time, then the probability of coalescence should have been relatively low during the time period immediately following the divergence of these populations, and higher during the time interval that immediately preceded this original divergence. In other words, the scenario of introgression from genetically divergent ancestral populations appears to predict a low coalescence rate during much of the last 700,000 years, and a high coalescence rate earlier in time. Inferences from the genomes of living Africans, from PSMC and other approaches, appear to show a contrasting pattern: a "wave" of effective population size over time, with an increasingly low effective population size (increased probability of coalescence) in the period leading from 300,000 back to 700,000 years ago, and an increasingly large estimated effective size (lower probability of coalescence) going further back in time.

Here I investigate whether the evidence from effective population size is consonant with evidence of introgression in human ancestral populations in Africa. I carry out numerical simulations to investigate how introgression and gene flow among genetically divergent populations affect the probability of coalescence at different times in the populations' histories. I also examine how the mutation process may affect data that emerge from a population history with introgression or migration between genetically divergent populations. The question is how much, if any, of the apparent evidence for population size changes in the prehistory of African populations may be explained by gene flow or introgression from genetically divergent ancestors.

## Methods

I consider here the case of independent, non-recombining genetic loci sampled from a single diploid genome in a population. This is convenient because of the relative simplicity of the simulation model and the comparability to PSMC approaches that examine evidence from single diploid genomes (Li and Durbin, 2011; Prüfer et al., 2013; Mallick et al., 2016). The purpose of the model is to examine in what way introgression and gene flow may affect the probability of coalescence in specific time intervals related to known simulated population histories. This question is of interest even in cases where gene flow was common enough that recombination may not have been substantially suppressed (in other words, where gene flow among genetically divergent populations might not be characterized as "introgression"), and therefore might not reject the hypothesis of random mating under common statistical approaches for detecting introgression (Plagnol and Wall, 2006). The model applied here is not sufficient to investigate the ancestral recombination graph of a population or to examine the power of inferences about gene flow or introgression.

The probability of coalescence in a random-mating population is expected to be a function of the population size. Here I deal only with constant population sizes, to specifically examine the effect of population structure. The assumption of constant population size is of course unrealistic as applied to African populations of ancestral humans; many of these populations have grown substantially during the last 50,000 years and possibly during much earlier time periods (Beltrame et al., 2016; Hsieh et al., 2016a; Mallick et al., 2016). These populations have also recently diversified in structure, with cross-coalescence among living African populations suggesting the existence of some aspects of today's population structure going back to as much as 200,000 years (Mallick et al., 2016). Recent population expansions would have reduced the probability of coalescence during the last interval of African population evolution, but aside from increasing the proportion of loci that coalesce prior to these expansions, these recent events would not change the relative probabilities of coalescence before any proposed archaic introgression. The issue of structure versus size as influences on the probability of coalescence is taken up in the Discussion.

**Population model.** The population model is illustrated in Figure 2. In the model, a single Wright-Fisher population $P_1$ of constant size is assumed to have existed from an indefinite time in the past up to the present day. At time $T_0$, this population instantaneously gave rise to a second population $P_2$. In each generation from $T_0$ up to a second time, $T_i$, a proportion $m$ of alleles in each population transferred to the other population, a symmetric island model of migration. At $T_i$, a fraction $q$ of the alleles in $P_1$ are derived by introgression from $P_2$. The model assumes this introgression did not change the population size of $P_1$, which is unrealistic but not too poor for small $q$. Time is scaled in terms of $2N$ generations, where $N$ is the population size of $P_1$ (which equals the population size of $P_2$). This is a similar model to that considered by Plagnol and Wall (2006) and later authors (Hammer et al., 2011; Hsieh et al., 2016b), with the differences that it does not incorporate any recent population expansions and it adds the possibility of migration between the genetically diverged populations prior to the time of introgression, instead of assuming complete reproductive isolation.

The model was implemented on Mathematica version 11.0, using the coalescent probabilities for a two-allele case with migration as described by Hudson (1990). The two alleles have a probability $1/2N$ of coalescing in each generation that they are in the same population, each has a probability $q$ of transferring to $P_2$ at time $T_i$, each has a probability $m$ of transferring from its population into the other population between $T_0$ and $T_i$, and any allele in $P_2$ at $T_0$ will transfer to $P_1$. Source code is available from Figshare. Each parameter was allowed to vary over a range of values in simulations. For each combination of parameter values, 50,000 genealogies were obtained, and the coalescence time (TMRCA) for the two alleles was recorded, in generations scaled to a factor of $2N$. While the probability of coalescence within a single population is given by $2N$, in the population model there are time increments where the effective probability of coalescence is less because of population structure. The effective probability of coalescence in a given time increment in the population model can be estimated as the ratio of the observed number of coalescence events in the time interval ($c_i$) relative to all genealogies that coalesce in this and earlier time intervals ($\sum_{j=i}^{\infty} c_j$). The effective population size in this time increment is then the reciprocal of the effective probability of coalescence.

**Mutation.** Genetic differences between alleles sampled in diploid humans reflect not only the genealogy of the alleles (modeled by the coalescent) but also the random process of mutation. Because mutations are rare events, the mutational process exerts substantial "noise" upon any attempt to estimate the TMRCA of a given locus. Sharp changes in the probability of coalescence will likely appear to be smeared across a relatively long period of time (Li and Durbin, 2011), and slight or short-term fluctuations might not be evident by examining the distribution of mutations. To examine the effects of the mutation process on the apparent TMRCA distribution, neutral mutations were added to the genealogies in a selected number of simulations. Under a constant rate of random neutral mutations, the expected number of mutations separating two alleles, E($H$), is a product of the neutral rate of mutations per site per generation, *mu*, the length of the locus in nucleotides, $L$, and 2 times the TMRCA in generations. In these simulations, the number of mutations for a given genealogy was modeled as a Poisson-distributed random variable with mean E($H$). The mutation rate, *mu*, was assumed to equal 3.5 times $10^{-8}$ per site per generation, and the length of each sampled locus was assumed to be 50,000 base pairs. Across 50,000 replicates, this roughly approximates the total length of a human genome.

## Results

A population model with introgression or migration between genetically divergent ancestral populations affects the distribution of TMRCA between two alleles by suppressing the effective probability of coalescence in the period when alleles may be resident in different genetically divergent populations. Figure 3 shows the TMRCA distribution in a population history with no introgression or migration, in which the effective probability of coalescence (Figure 3a) and the inferred effective population size (Figure 3b) are constant over time, with fluctuation only due to sampling. By comparison, a population history with introgression shows a sharp discontinuity in the effective probability of coalescence (Figure 3c) and inferred effective population size (Figure 3d) at the time of initial divergence of the genetically divergent ancestral populations.

Adding mutations to the genealogies results in a substantial smoothing of the history of inferred effective population size.  Figure 4 contrasts the inferred effective population size based upon the distribution of TMRCA from the coalescent (Figure 4a) with the inferred effective population size as estimated based on pairwise mutational differences between simulated 50-kb segments (Figure 4b). The sharp change in the effective probability of coalescence produced by the population model is smoothed substantially into a wave when mutations are added. Without modeling recombination, these results are not formally compatible with the history of effective population size for human genomes as inferred by PSMC. But the "wave" appearance of the population history as inferred from independent simulated loci is very similar in form to the wave that appears in PSMC-generated human population histories.

The results show very little difference when comparing sudden introgression versus slow, long-term gene flow. Both these processes result in a wave of similar form in the inferred history of effective population size (Figure 5). The key feature of the data in both cases is the initial divergence of populations at $T_0$ in the model. The transition from a random-mating ancestral population to two partially isolated populations separates two intervals with respectively high and low probabilities of coalescence, and the sharp transition is smoothed by adding mutations to the resulting genealogies. Whether subsequent migration between the populations is sudden or continuous, it has broadly similar results on the inferred history of effective population size.

The amount of introgression or gene flow from the genetically divergent ancestral population ($P_2$ in the population model considered here) determines the height of the wave of inferred effective population size. Figure 6 demonstrates the effect of different levels of introgression on the inferred population history. Human data are consistent with a 1.5 to 3-fold change in effective population size from the trough to the crest of the wave. This level of change in population size would require a rather large contribution of the genetically divergent ancestral population, for example, a level of introgression of 10% at 50,000 years ago combined with 0.2 migrants per generation (= $2Nm$) between $T_0$ and $T_i$ in the model. Alternatively, it is possible that populations involving introgression from multiple genetically divergent ancestors might explain the height of the wave in inferred effective population size.

The time intervals affected by the wave of effective population size are determined by the time $T_0$ in the population model considered here. Figure 7 shows the effect of this time of population divergence upon the wave, with successively older times of divergence giving rise to both older waves and waves of greater amplitude, all other parameters being held constant. Evidence of introgression from populations that diverged earlier in time is stronger than for populations that diverged more recently.

The time scale in all figures here is expressed in terms of $2N$ generations. If this scale were translated into terms of human population history, under the assumption that humans have a long-term $N$ = 10,000 and a generation length of 25 years, then a value of "1" on the time axis of each chart would be equivalent to 500,000 years ago (=20,000 generations), "2" would be equivalent to 1 million years ago, and "5" at the rightmost side of each graph is equivalent to 2.5 million years ago. Based on this relationship, it is possible to find the best-fit value of $T_0$ in comparison with published inferences of human effective population size history. One such population history is illustrated in Figure 7b. Again, under the assumption that $N = 10^4$ in the ancestral African population, introgression or gene flow from a genetically divergent ancestral population that diverged between 400,000 and 800,000 years ago would provide an acceptable match to PSMC inferences of human effective population size.

Human effective population histories inferred by PSMC and other approaches consistently show a larger effective population size in periods earlier than a million years ago. The models considered here do not consistently produce this aspect of published inferences of prehistoric human effective sizes. There are instances in which the inferred effective population size does appear higher in the earliest phases of the population history, for example, Figure 5b and Figure 6c. These are deviations from the expectation that the effective size will be 2$N$ prior to the divergence of genetically differentiated populations at $T_0$, which can be explained by random fluctuations among the small sample of genealogies that have TMRCA earlier than around 3$N$ generations. However, even such examples do not really appear to match the consistently larger effective size inferred for human populations prior to a million years ago.

## Discussion

Human populations had varied histories in the last 100,000 years, but in time intervals prior to 100,000 years ago, PSMC has generated very similar inferred histories of effective population size for different present-day populations. The results of this study suggest that some aspects of this consistent inferred population history can be explained by gene flow or introgression from genetically divergent ancestral populations. The wave of human population history during the period between 100,000 and 1 million years ago is best matched by introgression or gene flow from populations that diverged between 500,000 and 1 million years ago. This is a similar range of values as obtained by approaches that use other aspects of human genetic data to infer a history of introgression, not only in the ancestry of non-African populations (Sankararaman et al., 2014; Vernot and Akey et al., 2015), but in the ancestry of today's African populations (Plagnol and Wall, 2006; Hammer et al., 2011; Lachance et al., 2012; Hsieh et al., 2016b).

It is notable in the results that a very small fraction of introgression (on the order of 5% or less, Figure 4b, 6b and 6c) still gives rise to a pronounced wave in the inferred history of effective population size. Likewise, a very small amount of long-term gene flow also gives rise to a pronounced wave. Figure 5a, with $2Nm = 0.2$, uses a rate of gene flow less than one-twentieth the rate compatible with the current $F_{ST}$ of human global populations. How can a small fraction of introgression or gene flow have such a large effect on inferred ancestral effective population sizes? This may seem paradoxical considering that small amounts of introgression have a very small effect on genome-wide heterozygosity, which is the primary evidence for long-term effective population size. Introgression from Neandertals, for example, has not greatly increased the heterozygosity of non-African populations despite the great genetic divergence between Neandertal and modern genomes. But PSMC approaches go well beyond genome-wide heterozygosity to examine the distribution of heterozygosity across regions of the genome. It thus draws upon higher moments of a complex distribution, which is affected by gene flow and introgression in complex ways. Introgression may slightly decrease or increase the fraction of genetic loci that occur in a narrow window of heterozygosity values, when under neutrality only a small fraction of loci occur in that window to begin with. Hence, introgression or gene flow may have an outsized effect on the inference of effective population size for the time intervals that correspond to these differences.

Nothing about these results is inconsistent with the hypothesis that some large changes in actual population size may have occurred in the distant ancestry of human populations, irrespective of introgression or gene flow. In fact, under the population model considered here, the effective population size of the ancestral population *as a whole* is indeed larger during the time that ancestral lineages may exist within the genetically divergent second population. The inference of a larger effective size is accurate, it is simply explained by the presence of a ghost population. Ghost populations that are sources of introgression or gene flow are also ancestors of living human populations. The central point is that both population size and structure affect the relevant aspects of genetic variation, which means we cannot make progress understanding one demographic phenomenon without also considering the other.

PSMC examination of Neandertal and Denisovan genomes shows that their ancestral populations underwent a very different population history from the African ancestors of living human populations after they diverged (Prüfer et al., 2013). On the other hand, for the time intervals prior to the divergence of these archaic human populations, their inferred history of effective population size was congruent with the inferred history of living human populations (Figure 1b). The time that these histories come into congruence likely reflects the time of genetic divergence of these populations; it also approximates the minimum point in the wave of inferred human effective population size history. It is a good hypothesis that this minimum point in fact corresponds to the time immediately before the divergence of Neandertal and Denisovan from African human populations, when the probability of coalescence was highest between introgressed alleles in living non-Africans and the alleles that had been resident in the African ancestors of living non-Africans. In this case, the later part of the wave, representing higher inferred effective population size in human ancestral populations, is what may be explained by introgression or gene flow with genetically divergent archaic human groups. For present-day populations of non-Africans, this introgression came in part from Neandertals. Inside Africa, it may have been from other archaic human groups, with approximately the same inferred divergence date as Neandertals, as suggested by Lachance et al. (2012).

However, there are reasons to be skeptical of this scenario. The Neandertals and Denisovans share a common stem population, so the fact that they both have a similar history of divergence from the African ancestors of modern humans is not a chance coincidence. But it would be remarkable if one or more African archaic human populations mirrored precisely the same population history. Many different genetically divergent source populations for introgression may have existed within Africa, from archaic human populations as represented by the Kabwe, Florisbad, or Iwo Eleru crania, to "near-modern" human populations that nonetheless were strongly morphologically variable (reviewed by Stringer, 2016; Bräuer, 2008), possibly to highly-divergent hominin populations such as *Homo naledi* (Berger et al., 2015). Yet with all these candidates as possible sources for introgression, living populations in Africa have almost the same inferred history of effective population size as non-Africans across the period that these source populations diverged from each other. The similarity of inferred population histories for these different living groups of humans with different histories of introgression deserves more critical examination.

Likewise deserving of investigation is the inference of larger effective population size of human ancestral populations prior to 1 million years ago (Figure 1). The results from PSMC that point to a larger effective size in these distant ancestors are paralleled by approaches using rare insertion polymorphisms (Huff et al., 2010), suggesting a real phenomenon. It may be that this larger effective size reflects different population dynamics in the distant ancestors of humans, as may also be present in great ape species like chimpanzees, gorillas, and orangutans, which each have larger inferred effective population sizes than humans. Such dynamics may include introgression from genetically divergent populations of earlier hominins prior to 1 million years ago. This kind of introgression has already been documented for the Denisova 3 genome (Meyer et al., 2012; Prüfer et al., 2013). A number of morphologically diverse hominin populations inhabited Africa prior to 1 million years ago, and hybridization and introgression among them may have been an important part of the evolution of the genus *Homo*. This relatively remote interval of human population history is represented by only a small fraction of genetic loci across the genome, but may illuminate a key time in the origin of humans.

The discovery that introgression has contributed substantial variation into modern human populations has had great significance for the understanding of human variation (Hawks, 2013; Vattathil and Akey, 2015; Racimo et al., 2015). There is a growing recognition that this process of introgression or gene flow from genetically divergent ancestral populations played a role in the emergence of modern human populations in Africa (Beltrame et al., 2016; Ackermann et al., 2016). Considering that introgression or gene flow were widespread during human origins, the ancient divergence between archaic and modern human populations that interacted with each other may be one of the strongest influences on genetic diversity in humans today. The current study suggests that the effective population size inferred for particular intervals of time in the past is strongly influenced by the history of introgression or gene flow, even when the proportion of genetic variation derived from such introgression amounts only to a few percent of the ancestry of present-day people. This genetic contribution is very likely to have given rise to adaptive genetic variants that were valuable for modern human populations (Hawks and Cochran, 2006; Hawks et al., 2008; Vattathil and Akey, 2015; Ackermann et al., 2016). To the extent that such introgression or gene flow may also have occurred in earlier phases of human evolution, it was likely one of the key factors contributing to the success of human ancestors.

**Figure 1. History of human effective population size based upon PSMC of whole genomes.** In the top panel, the first PSMC inferences of human population history from Li and Durbin (2011) were based upon draft genomes from Yoruba, European ancestry, Korean, and Chinese individuals. The middle panel shows the results of PSMC analysis from Prüfer et al. (2013) including individuals from 11 diverse human populations and the Denisova and Altai Neandertal genomes. The bottom panel shows analyses from Mallick et al. (2016), including individuals from 3 African hunter-gatherer populations (Mbuti, Ju|'Hoan, and ‡Khomani San), 3 African agricultural and pastoral societies, and one French individual. The charts have different scales based on whether they leave data in a mutation time scale or attempt a conversion to years; all are presented with the *x*-axis (time dimension) as a logarithmic value. Mallick et al. (2016, bottom) did not report any inference for times earlier than 500,000 years. Despite the great variety evidenced in the inference of effective population size during the last 100,000 years of each population, every modern human genome gives substantially similar results for the time intervals prior to 100,000 years ago. All have a clear "wave" of larger inferred effective population size with a "crest" or maximum around 200,000 years ago, and a preceding "trough" or minimum around 500,000-800,000 years ago (beyond the timescale represented by Mallick et al., 2016).

**Figure 1 (continued).**

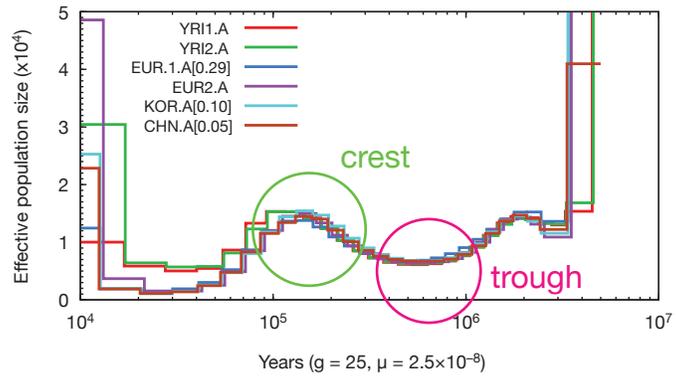

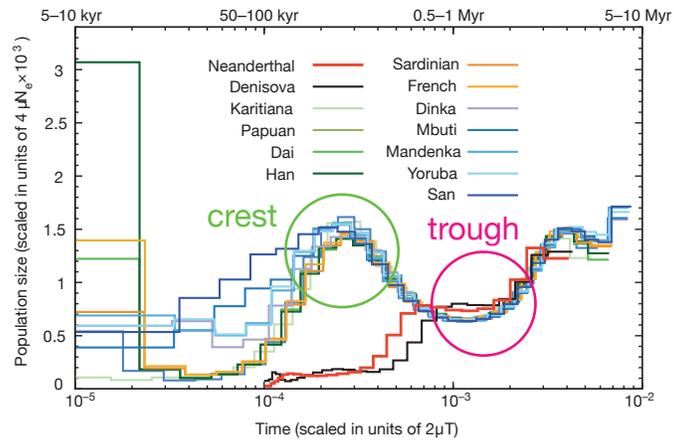

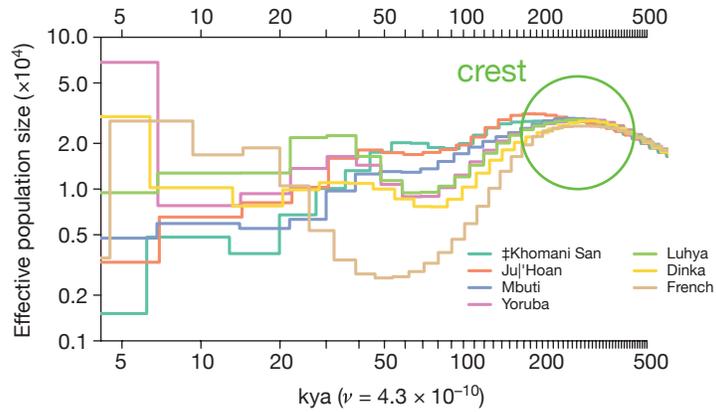

**Figure 2: Population model used in this study.** $P_1$ is a Wright-Fisher population of size $N$ that existed infinitely far back in time. $P_2$ came into existence instantaneously by diverging from $P_1$ at time $T_0$ in the past. From that time the two populations exchanged migrants at rate $m$, until time $T_i$, when a fraction $q$ of $P_1$ is derived by introgression from $P_2$.

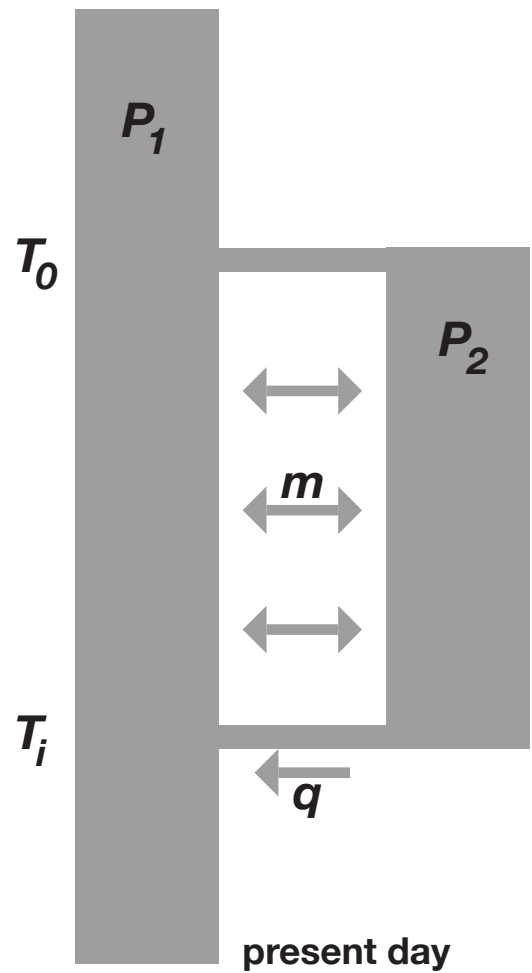

**Figure 3. Comparing population histories with introgression to the null model.** Panes (a) and (b) show the realized probability of coalescence and the inferred effective population size in a Wright-Fisher population model with no introgression or migration. Each data point represents the number of genealogies coalescing in a time increment of $0.05N$ generations. In (b), the effective size is estimated in each time increment, and the blue line represents a moving average of 10 time increments. Panes (c) and (d) show the same results for a population history in which 5% of the population derives from introgression $N$ generations in the past, from a second population that diverged $4N$ generations ago. There is an abrupt shift in the realized probability of coalescence corresponding to the divergence of the two populations $4N$ generations in the past. Points become more scattered moving toward the right (more ancient times) because the number of genealogies that coalesce in the most ancient time intervals is very small, lending sampling noise to the data.

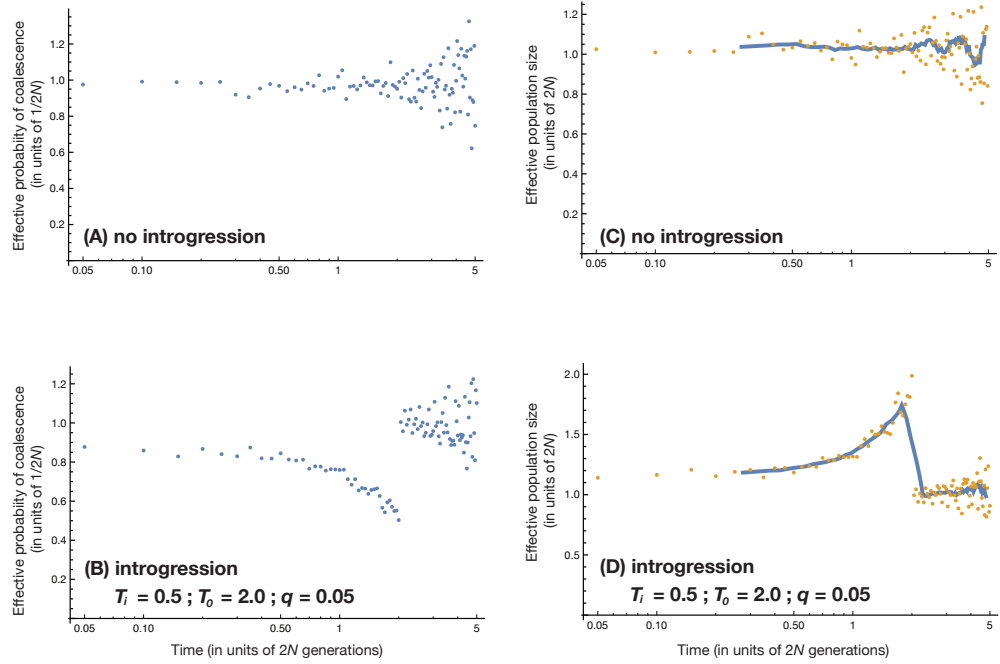

**Figure 4. Adding mutations to the genealogies turns a sharp transition into a wave.** (A) and (B) are based on the same simulation run, also illustrated in Figure 3c and 3d. (A) shows the effective population size for each interval as estimated from the fraction of loci with TMRCA in that interval; (B) shows the effective population size for each interval based upon the pairwise counts of mutations. In both (A) and (B), the blue line represents a moving average of the surrounding data points. Mutation adds a random component that tends to smooth sharp transitions in the dataset.

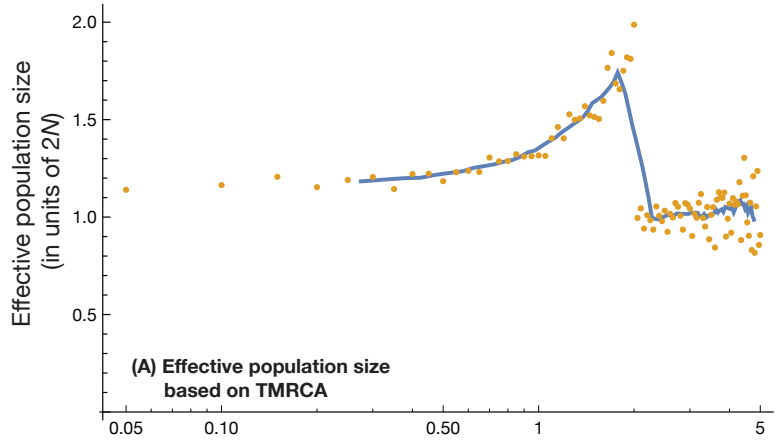

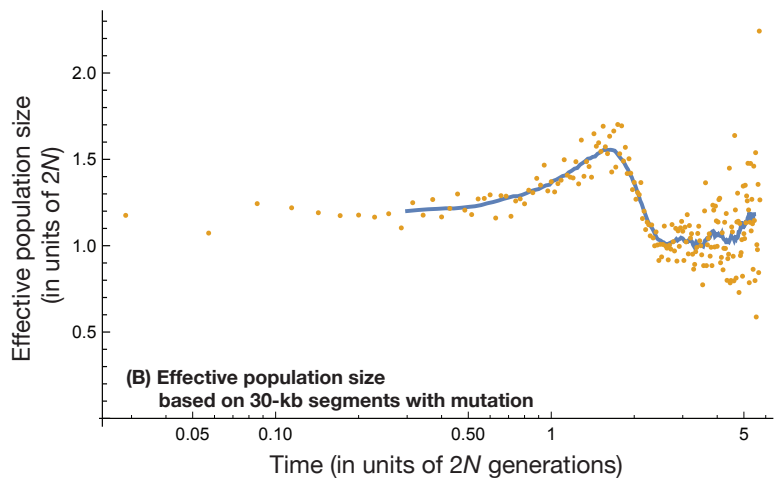

**Figure 5. Sudden introgression versus long-term gene flow.** In (A), 0.2 individuals per generation transfer between the two genetically divergent ancestral populations between $T_0$ and $T_i$, and there is no sudden introgression at $T_i$. In (B), there is no gene flow between the two populations, and 10% of population $P_1$ is the result of a sudden introgression from $P_2$ at time $T_i$. Both these scenarios result in very similar outcomes, with a wave of inferred effective population size at the same time and approximately the same magnitude. The upturn of inferred effective size in the rightmost (oldest) time intervals is not a reliable outcome of the simulations; this reflects the small sample of genealogies with the oldest coalescence times.

**Figure 5 (continued).**

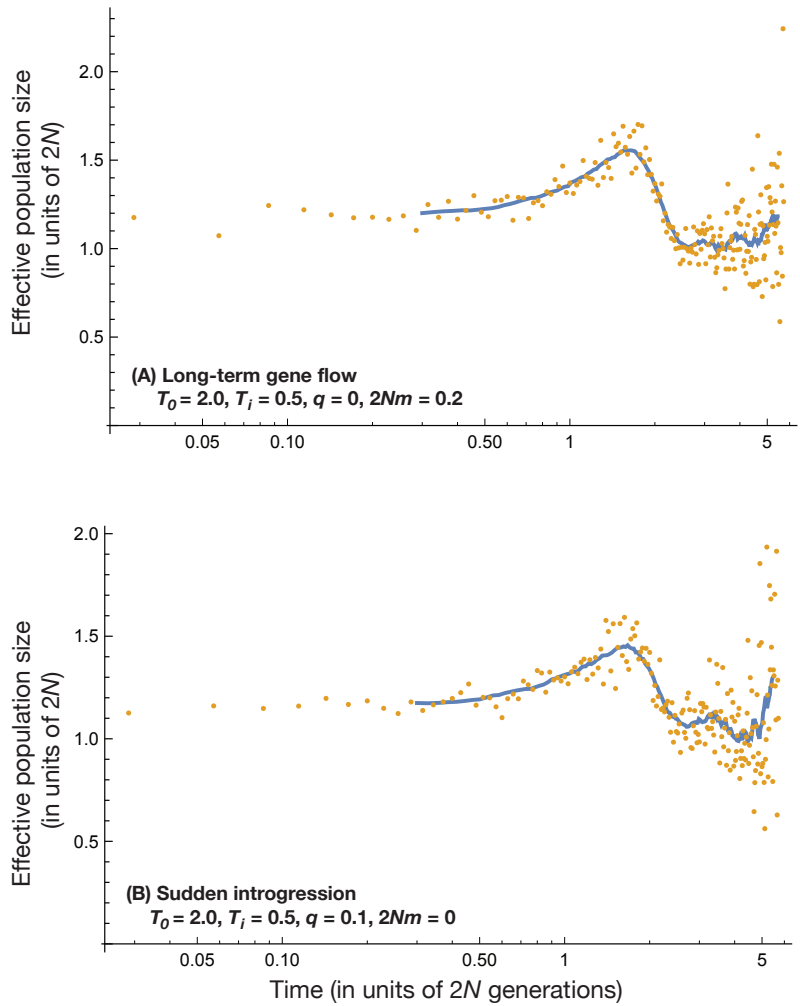

**Figure 6. Amplitude of the wave is a function of the amount of introgression.** Each panel represents an identical population model with introgression at 0.07 times 2*N* generations and divergence of $P_1$ and $P_2$ at 1.2 times 2*N* generations. In (A) 2% introgression; (B) 3% introgression; (C) 5% introgression; and (D) 8% introgression. The amplitude of the wave increases with greater introgression. If *N* = 10000 individuals and generations are 25 years long, then introgression in this model occurred approximately 35,000 years ago and the populations diverged approximately 600,000 years ago, roughly corresponding to some values estimated for African ancestral introgression.

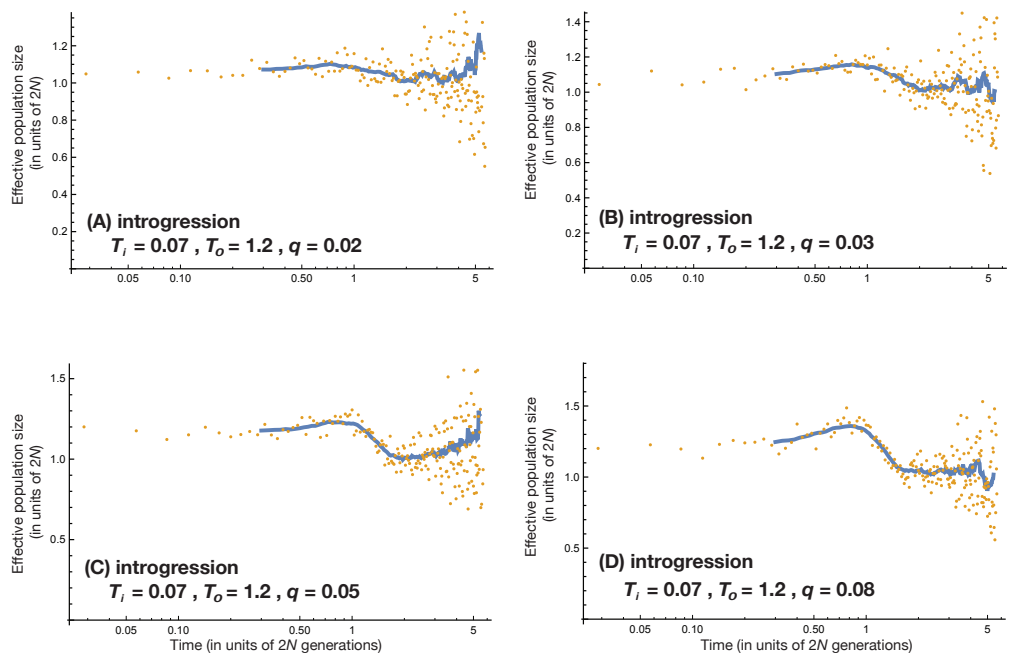

**Figure 7. Leading edge of the wave reflects time of divergence of introgression source population.** Each panel presents results with identical parameters except for the time of origin of the genetically divergent source population for introgression ($T_0$), which is indicated by the vertical red line. Migration ($2Nm$) in all panels is 0.1 individual per generation; the fraction of introgression ($q$) is 10%. Older time of origin also corresponds to greater wave amplitude, because the slight reduction in coalescence probability is cumulative over time in its effect.

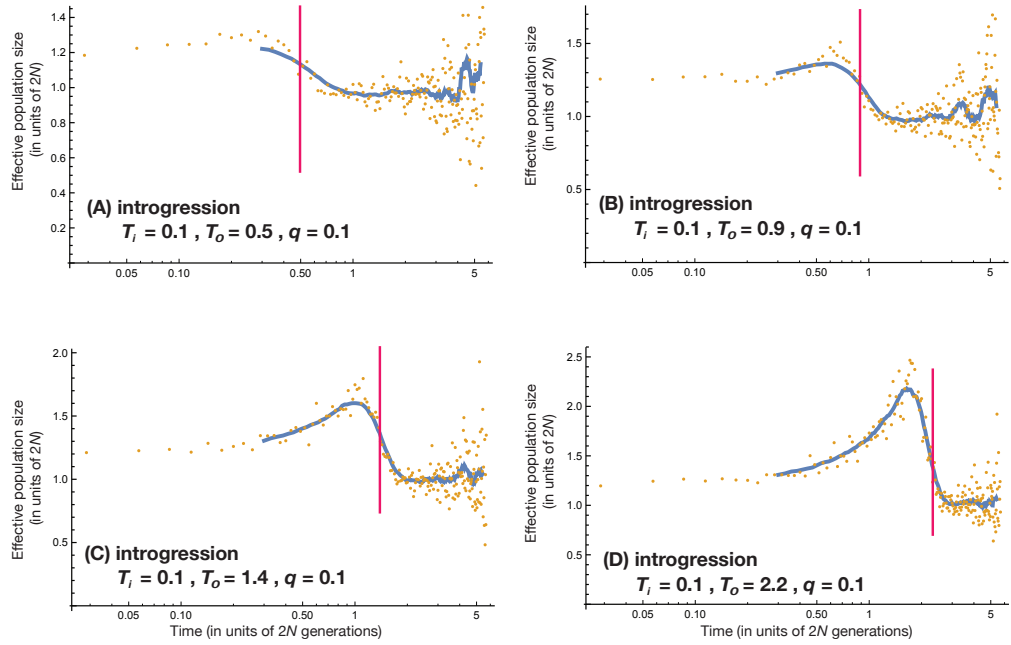